\documentclass[aps,prl,twocolumn,preprintnumbers,showpacs,showkeys]{revtex4}

\usepackage{graphicx}
\usepackage{amsfonts}
\usepackage{amsmath}
\usepackage{amssymb}
\usepackage{amsfonts}
\usepackage{graphicx}

\begin{document}
\bibliographystyle{unsrt}
\preprint{}
\title[noncollinear]{Identification of transverse spin currents in
noncollinear magnetic structures}
\author{Jianwei Zhang}
\affiliation{Department of Physics, 4 Washington Place, New York University, New York,
New York 10003}
\author{Peter M Levy}
\affiliation{Department of Physics, 4 Washington Place, New York University, New York,
New York 10003}
\author{Shufeng Zhang}
\affiliation{Department of Physics and Astronomy, University of Missouri-Columbia,
Columbia, Missouri 65211}
\author{Vladimir Antropov}
\affiliation{Ames Research Laboratory, Ames, Iowa 50011}

\begin{abstract}
We show that the transverse components of spin current in a ferromagnet is
linked to an off diagonal spin component of the transmission matrix at
interfaces;it has little to do with the mismatch of band structures between
dissimilar metals. When we take account of this component,not considered in
prior analyses, we find spin torque comes from a region of at lease $3$ nm
around an interface.
\end{abstract}

\volumeyear{year}
\volumenumber{number}
\issuenumber{number}
\eid{identifier}

\date{\today }
\published[Published ]{date}







\maketitle
Attention has recently been focused on current induced switching of magnetic
layers; this idea was proposed in 1996\cite{slon1} and has been verified in
the past three years.\cite{buhrman} It occurs when one drives an electric
current across a multilayered structure in which the magnetic layers are
noncollinear. The central idea behind the switching is that when the spin
polarized current that develops in a fixed ferromagnetic layer of a
multilayered structure impinges on a second noncollinear free magnetic layer
the component of the spin current transverse to the magnetization is
absorbed. As one assumes the conservation of spin angular momentum this
creates the torque on the background magnetization to switch it. One of the
unresolved issues is the length scale over which the transverse component of
the spin current is absorbed and what material parameters controls it. Here,
we point out, that the detail of the transverse spin near the interface does
matter in determining the spin transfer torque. The reason is that transport
is inherently non-local so that the incident spin polarization at an
interface does depend on the detail of the transverse spin accumulation
\textit{even if this length scale is small}, i.e., one must include
transverse pin accumulation. Mathematically, the boundary condition solely
from the vanishing transverse spin density in ferromagnets is not sufficient
to determine the spin torque. If we write the boundary condition to the
distribution function on the two sides of an interface, we have found that
the truncation of the transmission matrix to a diagonal matrix in spin space
misses key relations between the spinor distribution functions on the two
sides of an interface, and thus the transport across the entire structure is
unable to be self-consistently determined.

We recently found a possible mechanism by which one can inject off diagonal
spin distributions into each sheet of the spin split Fermi surface of a
magnetic layer, i.e., in the presence a current a spin flip potential can
exist at interfaces in noncollinear magnetic multilayers.\cite{levy} This
additional scattering excites the transverse components of spin currents in
magnetic layers that are found by adopting the Boltzmann equation's
definition of spin current and then we find there is no discontinuity in the
spin current at an interface between two dissimilar layers. This transverse
current is different from that evaluated from equilibrium states;\cite%
{stiles1,slon2} they \textit{complement} one another but the existence of
the latter does not negate the former. We evaluate the relevant parameters
over which this component is absorbed by using ab-initio band structure
calculations and f{}ind the characteristic length scale in a typical 3d
transition-metal ferromagnet(Co) is 3 nm. This is an order of magnitude
greater than that found in prior analysis which evaluated a quantum
mechanical expectation value of the current over the Fermi surface.\cite%
{stiles1,slon2} As the layers being switched are typically 2-3 nm this
differences is important as it alters one's estimates of the critical
current for switching.

The spin current density at a point $r$ can be defined as\cite{camblong}
\begin{equation}
\vec{j}_{\alpha \beta }(r)=\frac{e\hbar }{im}\left\langle \Psi _{\alpha
}^{\dagger }(r)\overset{\leftrightarrow }{\nabla _{r}}\Psi _{\beta
}(r)\right\rangle  \label{a'}
\end{equation}%
where the expectation value of the operator is taken over a nonequilibrium
distribution function and $\alpha ,\beta $ are spin indices. This
\textquotedblleft quantum mechanical\textquotedblright\ definition of spin
current Eq.(\ref{a'}) has been evaluated for a selected set of equilibrium
states so as to make a spin polarized current enter a ferromagnet at an
angle to the magnetization rather than self-consistently determing the
nonequilibrium distribution.\cite{stiles1,slon2} Based on this definition
the transverse component of the spin current has been shown to be carried by
the correlation between the states of opposite spin on the Fermi surface $%
k_{M}=k_{F\uparrow },k_{m}=k_{F\downarrow }$.\cite{stiles1} This
identification has lead one to conclude that it is the band mismatch between
the electronic structures of adjacent layers that causes the polarization of
the spin currents to change abruptly in a narrow region about an interfaces
so that the component of the spin current transverse to the magnetization of
a layer is absorbed within about a $\frac{1}{2}$ nm of the interface\cite%
{stiles1,slon2}, i.e., the spin split band structure of the magnetic layers
does not allow a spin current transverse to the magnetization to propagate
past the interfacial region.\cite{bauer,stiles2}
We recently showed how in the presence of a current
one can inject a spin distribution transverse to the magnetization across an
interface with a well defined momentum;\cite{levy} here we determine how it propagates
in a homogenous layer.

The equation of motion for the electron transport density matrix
$\left\langle a_{k\alpha}^{\dagger}a_{k^{\prime}\beta}\right\rangle $, i.e.,
the quantum or semiclassical Boltzmann equation, in a translationally
invariant system and for linear response is an equation for a given momentum
state in which states at other momenta enter through the scattering terms.%
\cite{smith} The Wigner transform of these matrix elements,\cite{smith}%
\begin{equation}
f_{\alpha\beta}(k,r)= \frac{e}{(2\pi\hbar)^3} {\displaystyle\int}
dk^{\prime}e^{ik^{\prime}\cdot r}\left\langle a_{k+k^{\prime}/2,\alpha
}^{\dagger}a_{k-k^{\prime}/2,\beta}\right\rangle  \label{b}
\end{equation}
is independent of position $r$ for a system in equilibrium, e.g., the Fermi
distribution function. While the perturbation of the electric field in and
of itself does not break the translational invariance of a layer the ensuing
current crossing an interface does; therefore when a current is driven
across a system as a multilayer the out of equilibrium part of $%
f_{\alpha\beta}(k,r)$ develops a position dependence. In the
Boltzmann description the transverse components of the spin
current are
\begin{equation}
\vec{j}_{\alpha\beta}(r)=\frac{1}{(2\pi\hbar)^3} \int dk\vec{v}%
(k)f_{\alpha\beta}(k,r),  \label{c}
\end{equation}
where in linear response $\vec{v}(k)$ is the Fermi velocity of the electron.
In principle both definitions Eqs.(\ref{a'}) and (\ref{c}) give the same
current provided sufficient care is taken in their evaluation.

The equation of motion for the density matrices or distribution
functions is the Boltzmann equation; here we use the semiclassical
version of this equation. In each layer the energy and density
vary slowly on the length scale of the Fermi wavelength so that we
can limit ourselves to the first term in the gradient expansion of
the equation of motion.\cite{rammer and smith} When we limit
ourselves to linear response in the external electric field we
f{}ind the equations of motion \textit{in steady state} for the
elements of the spinor density matrix for each momentum state on
the Fermi surface $k_{p}$ are,\cite{smith}
\begin{equation}
v_{p}^{x}\partial_{x}f_{p}-eEv_{p}\delta(\varepsilon-\varepsilon _{F})=-
\frac{f_{p}-\left\langle f_{p}\right\rangle}{\tau_{p}} -\frac{%
f_{p}-\left\langle f_{p^{\prime}} \right\rangle}{\tau_{sf}},  \label{d}
\end{equation}
and
\begin{equation}
v_{p}^{x}\partial_{x}f_{p}^{\pm}\mp i\frac{J_{p}}{\hbar}f_{p}^{\pm}=- \frac{%
f_{p}^{\pm}-\left\langle f_{p}^{\pm}\right\rangle}{\tau_{p}},  \label{e}
\end{equation}
where we have used a simplified index $p$ to denote the momentum $k_{p}$ of
a state on the $n^{th}$ sheet of the Fermi surface(we suppress this index), $%
p^{\prime}$ are states of opposite spin to $p$, the average $\left\langle
f_{p}\right\rangle $ represents elastic scattering to all states on the
Fermi surface, $v_{p}^{x}$ is the component of the Fermi velocity along the
electric field $E$, $J$ is the magnetic part of the energy (see below) and
we have limited the current induced variations of the distribution function
to those along the growth direction of a multilayered structure $x$ (also
the field direction). The diagonal elements $f_{p}=f_{\alpha\alpha}(k_{p},x)$
represents the occupancy of the state $k_{p}$; in equilibrium it is given by
the Fermi function so that only the spin state $\alpha$ that crosses the
Fermi level is occupied while the other is zero and we do not consider it
further. The off diagonal elements $f_{p}^{\pm}\sim
f_{\uparrow\downarrow}(k_{p},x)$, which we call a current induced spin
coherence,\cite{coherence} represent coherences between the state $k_{p}$ on
the Fermi surface and the states with opposite spin; see Eq.(\ref{b}); these
coherences occur when we drive a spin current across a normal
metal/ferromagnet (N/F) interface.\cite{levy} The scattering terms include
those for states of the same spin on the Fermi surface $\tau_{\alpha}^{-1}$
as well as those between sheets of opposite spin $\tau_{sf}^{-1}$; as $%
\tau_{sf}^{-1}\ll$ $\tau_{\alpha}^{-1}$ we include the latter only to have
well defined boundary conditions on our distribution functions.

The solutions for the longitudinal components are well know.\cite{valet-fert}
From Eqs.(\ref{d}) and (\ref{e}) we see that the electric field only
creates out of equilibrium longitudinal components of the distribution
functions; in a homogeneous magnetic layer there is no coupling to the
transverse components. However when the spin current from one layer is
injected into another noncollinear magnetic layer the transverse components
can be excited.
As such they have to be determined by a self consistent calculation of the
distribution functions in the different layers of a multilayered structure.
When we neglect collisions (the rhs of Eq.(\ref{e})) the transverse
solutions are $f_{p}^{\pm }(x)\sim \exp \pm \mathbf{i}(J_{p}/\hbar
v_{p}^{x})x$; when we average this over the Fermi surface we f{}ind an
interference between individual $p$ or $\mathbf{k}$ states so that the
transverse components of the spin currents in the magnetic layers, $%
j_{x}^{\pm }(x)\sim \int v_{x}(k)f^{\pm }(k,x)dk$ (Eq.(\ref{c})), can be
f{}it to a form \textit{approximating} an exponential decay $\sim \exp
-x/\lambda _{tr}$.\cite{stiles power law decay} In the ballistic regime $%
\lambda _{tr}=d_{J}\equiv h\overline{v_{F}/J}$ where the bar denotes an
average over states on the $n^{th}$ sheet of the Fermi surface under
consideration;\cite{jianwei} while for diffusive systems where we consider
the collision terms $\lambda _{tr}=\lambda _{J}\equiv \sqrt{d_{J}\lambda
_{mfp}/3\pi }$.\cite{prl} Therefore when components of the spin current are
injected into a magnetic layer that are transverse to its magnetization, we
find they propagate a distance $\lambda _{tr}$ before decaying; as this
distance is an order of magnitude greater than the Fermi wavelength one can
describe the transverse spin currents in the semiclassical Boltzmann
approach. As $\lambda _{tr}$ is comparable to the thickness of the magnetic
layers undergoing switching one cannot assume the transverse components of
spin currents are entirely absorbed in such thin layers.

To evaluate $d_{J}$ we need the Fermi velocity $v_{F}(\mathbf{k)}$, and
$J(\mathbf{k})$ for each point $\mathbf{k}$ on the $n^{th}$ sheet of the Fermi
surface under consideration; the latter is identified as follows. The energy
of a spin polarized band in a uniform ferromagnetic metal can be written as
a spinor
\begin{equation}
\hat{\varepsilon}_{n}(k)=\varepsilon _{n}(k)\mathbf{1}+\frac{1}{2}J_{n}(k)%
\vec{M}\cdot \vec{\sigma}  \label{f}
\end{equation}%
where $n$ is the band index, $\vec{M}$ is a unit vector in the direction of
the magnetization, and $\vec{\sigma}$ represents the Pauli spin matrices. As
such $J_{n}(k)$ is the magnetic part of the band energy. For a band crossing
the Fermi surface it is found by identifying the band of opposite spin that
is split from the common energy $\varepsilon _{n}(k)$ by $J_{n}(k)$. This
requires one to do two band structure calculations, one nonmagnetic in which
each band is spin degenerate and another for the fully spin polarized state
in which the two spin split states originating from the common degenerate
band can be easily identified along directions of high symmetry.\cite{ingrid}
With this identification we find%
\begin{equation}
J_{n}(k)=\left( \varepsilon_{n\uparrow}(k)-\varepsilon_{n\downarrow}
(k)\right),  \label{g}
\end{equation}%
where $\varepsilon _{n\sigma }(k)$ are the energies of the spin split band
and $n$ is the index for the band that crosses the Fermi surface whose
magnetic component we are evaluating. Therefore $J_{n}(k)$ can be found from
looking at the band structure of a ferromagnetic metal; it is the difference
in energy at constant $k$ \ between two split bands. While these two bands
mostly lie on top of (next to) one another, there are regions of the
Brillouin zone where one has to do some further analysis in order to
identify the bands which had a common origin in the nonmagnetic phase.

From the band structure for $fcc$ Co we f{}ind $d_{J}$ for $\Delta_{1}$
majority states along the $\Gamma-X$ (100) direction is 7.4 nm but it
decreases rapidly off this axis, along $\Gamma-K$ (110) $d_{J}$ is 2.93 nm,
along $\Gamma-L$ (111) $d_{J}$ is 1.1nm, and the average $d_{J}$ for the
majority band over the Fermi surface is about 3 nm.\cite{tail} Of the three
minority spin bands crossing the Fermi surface the one that is the partner
of the majority, \#3, has a $d_{J}$ of 0.8 nm while the other two have even
shorter $d_{J}$. Central to the semiclassical description of the transverse
spin currents is the condition that they can propagate over distances large
compared to the Fermi wavelength; we find that this condition is well
satisfied for the 3d transition metal ferromagnets for the majority band.

To identify the source of the discontinuity in a spin current at interfaces
we have solved the Boltzmann equations Eqs.(\ref{d}) and (\ref{e}) for the
current across two magnetic layers; while there is an intervening
nonmagnetic layer we adopt the usual assumption that its thickness is much
less than the spin diffusion length so the spin current is constant across
it. For the electronic structure we take a free electron-like model in which
the degenerate spin bands are split by the internal fields; the effect of
the exchange splitting of the bands is to produce two separate channels of
conduction on the Fermi surface, one majority one minority.\cite{bands} To
keep track of current induced coherences the conduction in each channel is
described by a $2\times 2$ density matrix so that one has to match \textit{%
two} $2\times 2$ density matrices at interfaces to obtain a description of
the spin current across a noncollinear multilayered structure, i.e.,
\textit{each} spin channel is represented by a spinor distribution function. To
perform a self consistent calculation of the distribution functions across a
multilayered structure we have to match them at the interfaces by using
transmission and reflection coefficients. In addition to the spin diagonal
coefficients $T_{mm\Longrightarrow ss^{\prime}}(s^{\prime}=s)$ previously
considered we add the off diagonal coefficient with $s^{\prime}\neq s$ that
transmits spin coherence across an N/F interface;\cite{levy} therefore we
must consider in our matching not only states on the Fermi surface but also
those with which electrons form current induced coherences in steady state,
i.e., in general we have to match $4\times 4$ rather than $2\times 2$
distribution functions across interfaces.

When one limits oneself to spin diagonal transmission coefficients at an
interface there is no transmission into off diagonal or transverse elements
of the density matrix, i.e., the $f_{\uparrow \downarrow }$, are zero (not
excited) in the magnetic layers; as they have been identified as the
transverse components of the spin current one arrives at the conclusion that
the transverse component of the spin current has been absorbed at the
interface.\cite{stiles1,bauer} However, when we include the off diagonal
transmission coefficient the discontinuity of the transverse component of
the spin current at interfaces is removed. In other words without $%
T_{mm\Longrightarrow ss^{\prime }}$with $s^{\prime }\neq s$ we find the spin
current in a ferromagnetic layer is parallel to the background magnetization
and there is an abrupt rotation of the spin current when the layers are
noncollinear; with it the spin current is continuous. While transverse spin
currents do not exist when one uses the equilibrium distribution that was
selected to evaluate Eq.(\ref{a'}), \cite{stiles1,slon2} they \textit{do}
exist when we use Eq.(\ref{e}) to self-consistently evaluate the transverse
component of the spin current Eq.(\ref{c}) in steady state.\cite{steady}
When we solve the Boltzmann for the spin current across two noncollinear
magnetic layers we f{}ind its polarization starts to deviate from the
background magnetization a distance $\lambda _{tr}$ on either side of the
interface and is continuous across it.\cite{future} While the spin
accumulation, and therefore the resistance, depends on the ref{}lection and
transmission coefficients due to the specular scattering at the interface,
the continuity of the spin current is insensitive to them as long as we
include $T_{mm\Longrightarrow ss^{\prime }}$with $s^{\prime }\neq s$ ;
therefore we conclude that their discontinuity is not an indigenous property
of noncollinear magnetic layers. We f{}ind it is caused when one suppresses
the transverse components of the spin currents in the magnetic layers either
by using only diagonal transmission coefficients \cite{bauer} or by using
only states at the Fermi surface to evaluate the spin current Eq.(\ref{a'}).%
\cite{stiles1,slon2}

In summary, when the \textit{steady state} spin current is self-consistently evaluated
using Eq.(\ref{c}) we keep track of the transverse spin accumulation around
the interfaces irrespective of its length scale.\emph{\ }We find there is no
absorption at the interface, it is continuous across it, and the transverse
component decays within 3 nm, at least for Co, irrespective of the
reflection and transmission coefficients; albeit these coefficients affect
the magnitude of the spin currents and resulting torques. Therefore the
discontinuity in spin currents at interfaces is not due to band mismatch;
rather we find it arises from omitting the off diagonal components of the
transmission coefficients and the ensuing transverse spin accumulation. The
difference between our result and those which predict spin torques created
within $\frac{1}{2}$
nm of  an interfaces is important as the thickness of layers undergoing
magnetization reversal is in the range of $2-3$ nm. When we introduce diffuse
scattering in the layers we reduce the decay length for the transverse
component of the spin current as we found in the spin diffusion model of
current induced magnetic switching,\cite{asya} but we do not create
discontinuities in the current.

\begin{acknowledgments}
We thank Albert Fert for questioning the validity of the spin diffusion
model in this context, Ingrid Mertig for identifying the magnetic component
of the spinor band energy from ab-initio calculations and together with
Peter Zahn for the band structure of Co and the Fermi velocity across the
various sheets of the Fermi surface, and Asya Shpiro for comparing her
results from the spin diffusion model to those presented here. Numerous
discussions with Andy Kent, Jerry Percus, Bob Richardson and Tycho Sleator
at NYU were very useful especially in developing the relevance of coherences
in the context of transverse spin currents. This work done under a grant
from the National Science Foundation (Grant DMR 0131883 and DMR 0314456).
Ames Laboratory is operated for the U.S.Department of Energy by Iowa State
University under Contract No. W-7405-82.
\end{acknowledgments}


\begin{thebibliography}{99}
\bibitem{slon1} J.C. Slonczewski, J. Mag. Mag. Mater. \textbf{159}, L1
(1996); \textit{ibid} \textbf{195}, L261 (1999); L. Berger, Phys. Rev. B
\textbf{54}, 9353 (1996); J. Appl. Phys. \textbf{89}, 5521 (2001).

\bibitem{buhrman} J. A. Katine, F. J. Albert, R. A. Buhrman, E. B. Myers,
and D. C. Ralph, Phys. Rev. Lett. \textbf{84}, 3149 (2000); F. J. Albert, J.
A. Katine, R. A. Buhrman, and D. C. Ralph, Appl. Phys. Lett. \textbf{77},
3809 (2000); J. Grollier, V. Cros, A. Hamzic, J. M. George, H. Jaffr\`{e}s,
A. Fert, G. Faini, J. Ben Youssef, and H. Legall, Appl. Phys. Lett. \textbf{%
78}, 3663 (2001).

\bibitem{levy} Peter M. Levy and Jianwei Zhang, to be published.

\bibitem{stiles1} M.D. Stiles and A. Zangwill, Phys. Rev. B \textbf{66},
014407 (2002).

\bibitem{slon2} J.C. Slonczewski, J. Mag. Mag. Mater.\textit{\ }\textbf{247}%
, 324 (2002).

\bibitem{camblong} H.E. Camblong, P.M. Levy and S. Zhang , Phys. Rev. B%
\textbf{51}, 16052 (1995).

\bibitem{bauer} A. Brataas, Yu.V. Nazarov, and G.E.W. Bauer, Phys. Rev.
Lett. \textbf{84}, 2481 (2000) and D.H. Hernando, Y.V. Nazarov, A. Brataas,
and G.E.W. Bauer, Phys. Rev.B \textbf{62}, 5700 (2000); Alexey Kovalev, Arne
Brataas and Gerrit E.W. Bauer,\textit{\ ibid} \textbf{66}, 224424 (2002);
Gerrit.E.W. Bauer, Yaroslav Tserkovnyak, Daniel Huertas-Hernando and Arne
Brataas, \textit{ibid} \textbf{67}, 094421 (2003).

\bibitem{stiles2} M.D. Stiles and A. Zangwill, J.Appl. Phys.\textbf{\ 91},
6812 (2002).

\bibitem{coherence} The off diagonal elements of a density matrix $|\phi
_{a}><\phi_{b}|$ refer to coherences between states; only the diagonal
elements $a=b$ represent the occupancy of a state.

\bibitem{smith} see \textbf{Transport Phenomena} by H. Smith and H.H.
Jensen; in particular Sec. 1.13.

\bibitem{rammer and smith} J. Rammer and H. Smith, \ Rev. Mod. Phys. \textbf{%
58}, 323 (1986).

\bibitem{valet-fert} T. Valet and A. Fert, Phys. Rev. B\textbf{48}, 7099
(1993).

\bibitem{stiles power law decay} See Stiles Ref. 3 for an excellent
discussion of how oscillating but non-decaying functions can \textit{%
interfere} to produce a seemingly exponential decay.

\bibitem{jianwei} Jianwei Zhang, P.M. Levy and Shufeng Zhang, Bull. Amer.
Phys. Soc. \textbf{48}, 117 (2003); P.M. Levy, Jianwei Zhang and Vladimir P
Antropov, \textit{ibid }\textbf{48}, 821 (2003).

\bibitem{prl} S.\ Zhang, P.M.\ Levy and A.\ Fert, Phys. Rev. Lett. \textbf{88%
}, 236601 (2002).

\bibitem{ingrid} This identification of the magnetic part of the energy of a
band state has been suggested by Professor Ingrid Mertig, privgate
communication.

\bibitem{tail} The sharp peak in the distribution of $d_{J}(k)$ in the $%
\Gamma -X$ direction produces a long range tail in the transverse spin
current whose amplitude is small compared to the main response coming form
the average \ $\overline{d_{J}(k)}$. \ In the ballistic limit\ it does not
decay; however it will when one considers diffusive scattering in the bulk
of the layers.

\bibitem{bands} In Co for example there are three minority bands crossing
the Fermi surface. We consider the only one deemed to make a significant
contribution to electron transport.

\bibitem{steady} When the current is first turned on there is a
discontinuity in the transverse component of the spin current as envisaged
by others; however in steady state this is replaced by a more gradual
rotation of the polarization. Therefore in steady state the spin torques and
angular dependence of the resistance are different than those found from
calculations which do not take account of these accumulations.

\bibitem{future} The results of our self-consistent determination of
transport across noncollinear multilayers will be presented in a forthcoming
publication.

\bibitem{asya} A. Shpiro, P.M. Levy and S. Zhang, Phys. Rev.B \textbf{67}%
,104430 (2003).
\end{thebibliography}

\end{document}